\documentclass[runningheads]{llncs}
\usepackage{graphicx}
\usepackage{graphicx}
\usepackage{amsmath}
\usepackage{booktabs}
\usepackage{tabularx}
\usepackage{color}
\usepackage{multirow}
\usepackage{booktabs}

\begin{document}

\newcommand \ourmethod {MT-UDA}
\title{MT-UDA: Towards Unsupervised Cross-Modality Medical Image Segmentation with Limited Source Labels}
\titlerunning{MT-UDA}
%
\author{Ziyuan Zhao\inst{1, 2} \and Kaixin Xu\inst{2} \and
Shumeng Li\inst{1, 2} \and Zeng Zeng\inst{2} \and Cuntai Guan\inst{1}}

\authorrunning{Zhao et al.}
%
\institute{Nanyang Technological University, Singapore \and
Institute for Infocomm Research, A*STAR, Singapore
}
\maketitle              
\begin{abstract}
The success of deep convolutional neural networks (DCNNs) benefits from high volumes of annotated data. 
However, annotating medical images is laborious, expensive, and requires human expertise, which induces the label scarcity problem. 
Especially When encountering the domain shift, the problem becomes more serious.
Although deep unsupervised domain adaptation (UDA) can leverage well-established source domain annotations and abundant target domain data to facilitate cross-modality image segmentation and also mitigate the label paucity problem on the target domain, the conventional UDA methods suffer from severe performance degradation when source domain annotations are scarce. 
In this paper, we explore a challenging UDA setting - limited source domain annotations. 
We aim to investigate how to efficiently leverage unlabeled data from the source and target domains with limited source annotations for cross-modality image segmentation.
To achieve this, we propose a new label-efficient UDA framework, termed MT-UDA, in which the student model trained with limited source labels learns from unlabeled data of both domains by two teacher models respectively in a semi-supervised manner.
More specifically, the student model not only distills the intra-domain semantic knowledge by encouraging prediction consistency but also exploits the inter-domain anatomical information by enforcing structural consistency.
Consequently, the student model can effectively integrate the underlying knowledge beneath available data resources to mitigate the impact of source label scarcity and yield improved cross-modality segmentation performance.
We evaluate our method on MM-WHS $2017$ dataset and demonstrate that our approach outperforms the state-of-the-art methods by a large margin under the source-label scarcity scenario.

\end{abstract}
\section{Introduction}
Deep convolutional neural networks (DCNNs) have obtained promising performance on medical image segmentation tasks~\cite{long2015fully,ronneberger2015u}, which further promotes the development of automated medical image analysis.
DCNNs are data-hungry and require large amounts of well-annotated data, however, in real-world clinical settings, medical image annotations are pricey and labor-intensive, which require extensive domain knowledge from biomedical experts. 
This leads to that scarce annotations are available for training DCNNs,~\emph{i.e.}, label scarcity.

To alleviate the burden on human annotation, plenty of methods beyond supervised learning have been proposed for improving label efficiency on medical imaging~\cite{tajbakhsh2020embracing}, including self-supervised learning~\cite{zeng}, semi-supervised learning~\cite{bai2017semi,zhao2021dsal} and disentangled representation learning~\cite{chartsias2019disentangled}.
In recent years, semi-supervised learning (SSL) methods based on the self-ensembling strategy~\cite{tepens2017,meanteacher,cheplygina2019not} have received much attention in medical image analysis, achieving state-of-the-art results in many SSL benchmarks. 
For instance, Laine and Aila~\cite{tepens2017} propose Temporal Ensembling to enforce the consistent outputs of the network-in-training across different epochs. 
Tarvainen and Valpola~\cite{meanteacher} build the mean teacher (MT) model based on the exponential moving average (EMA) of the weights of the student network, forcing the prediction consistency and further boosting the model performance. 
Subsequently, many studies have been devoted to leveraging abundant unlabeled data based on MT to mitigate the paucity-of-label problem in biomedical image segmentation~\cite{cui2019semi,dualteacher,yu2019uncertainty}.
These methods, however, are presented for label-efficient learning on a single partially labeled dataset, failing to use cross-domain information well when using multi-domain datasets.

On the other hand, given the various imaging modalities with different physical principles, such as CT and MR, the domain shift problem is severe in cross-modality image segmentation, resulting in significantly reduced performance when applying well-trained DCNNs on one domain (\emph{e.g.}, MR) to another domain (\emph{e.g.}, CT), especially in the absence of target labels. 
To tackle this serious issue, much research has been devoted to investigating unsupervised domain adaptation~(UDA) for minimizing the discrepancy between the source and target domains, consequently boosting the generalization ability on the target domain for cross-modality medical image segmentation~\cite{ganin2015unsupervised,yang2019unsupervised}. 
Inspired by the great success of generative adversarial networks (GANs) on image-to-image translation~\cite{arjovsky2017wasserstein,GAN,hoffman2018cycada}, many approaches have been developed with adversarial learning from different perspectives for domain alignment, including image-level adaptation~\cite{chen2018semantic,zhang2018task}, feature-level adaptation~\cite{pnp_1,dou2018pnp,tzeng2017adversarial} and their mixtures~\cite{chen2019synergistic,chen2020unsupervised}. 
For example, Chen~\emph{et al.} design a synergistic image and feature adaptation model~\cite{chen2020unsupervised}, which achieves the state-of-the-art performance in UDA for cross-modality medical image segmentation. Despite the success of adversarial learning in UDA, these methods heavily rely on abundant source labels, which become sub-optimal when only limited source labels are available in clinical deployment.

These motivate us to advocate studying a practical, challenging, and different UDA setting from the past, where only limited source labels are accessible. 
In this paper, we investigate the feasibility of integrating SSL into UDA under source label scarcity and propose a novel label-efficient UDA framework for cross-modality medical image segmentation. 
We first present a dual cycle alignment module (DCAM) to bridge the appearance gap across domains, synthesizing \emph{source-like domain} images and \emph{target-like domain} images via adversarial learning~\cite{GAN}. 
We further develop an MT framework~\cite{meanteacher} for UDA, named \ourmethod, to exploit the knowledge from both intermediate domains. 
In~\ourmethod, the student model distills the intra-domain semantic knowledge by encouraging the prediction consistency of the source domain and exploits the inter-domain anatomical information by enforcing the structural consistency across domains. 
We evaluate the proposed \ourmethod~on a public multi-modality cardiac image segmentation dataset, MM-WHS 2017, and demonstrate that our method outperforms the state-of-the-art methods by a lot under the challenging UDA scenario.

\begin{figure}[t]
    \centering
    \includegraphics[width = 0.85 \columnwidth]{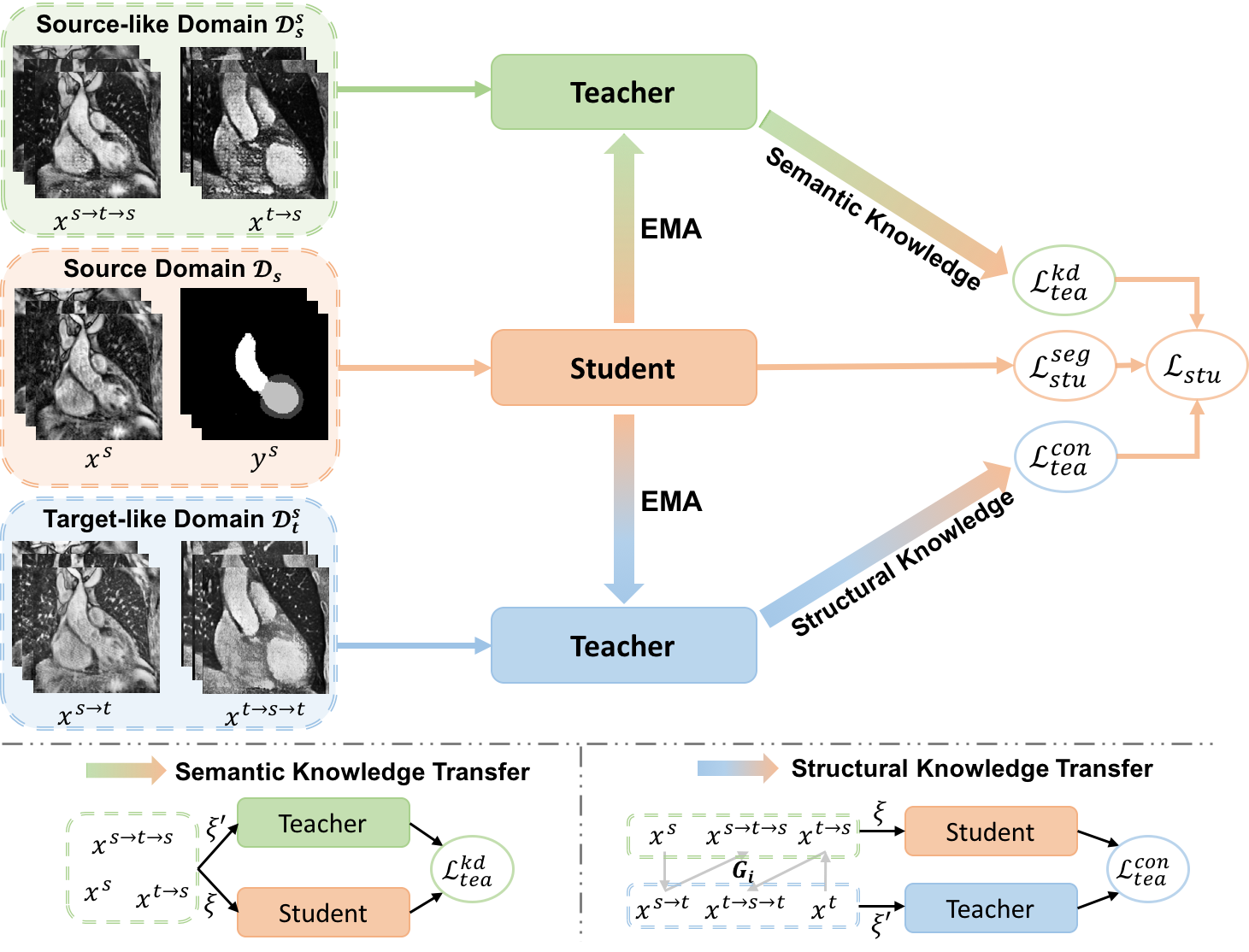}
    \caption{Overall framework of our proposed \ourmethod. The student model learns from labeled source samples $\mathcal{D}_{s}^l$ by the $\mathcal{L}_{{stu}}^{s e g}$ loss, and distills the intra-domain semantic knowledge and inter-domain anatomical information from \emph{source-like domain} and \emph{target-like domain} by $\mathcal{L}_{{tea }}^{{kd}}$ and $\mathcal{L}_{{tea }}^{{con}}$, simultaneously.}
    \label{fig:pipeline}
\end{figure}

\section{Methodology}

Let $\mathcal{D}_{s}^l=\left\{\left(\mathbf{x}_{i}^{s}, y_{i}^{s}\right)\right\}_{i=1}^{N}$ and $\mathcal{D}_{s}^u=\left\{\left(\mathbf{x}_{i}^{s}\right)\right\}_{i=N+1}^{M}$ denote the labeled samples and unlabeled samples from source domain (\emph{e.g.}, MR), respectively. 
In conventional UDA setting, abundant labeled source data $\mathcal{D}_{s}^l$ is given, \emph{i.e.}, $N = M$. Differently, in our setting, only limited labeled source data $\mathcal{D}_{s}^l$ is used for UDA, \emph{i.e.}, $N << M$, which is more practical and challenging. 
We aim to exploit $\mathcal{D}_{s}^l$, $\mathcal{D}_{s}^u$ and unlabeled samples $\mathcal{D}_{t}=\left\{\left(\mathbf{x}_{i}^{t}\right)\right\}_{i=1}^{P}$from target domain (\emph{e.g.}, CT) for UDA to improve the model performance on the target domain. 
The overview of the proposed method is presented in Fig.~\ref{fig:pipeline}. 
Firstly, two sets of synthetic images, \emph{i.e.}, \emph{source-like domain} $\mathcal{D}_{s}^s$ and \emph{target-like domain}, $\mathcal{D}_{t}^s$ are generated with the proposed dual cycle alignment module to alleviate the notorious domain discrepancy in appearance (see Fig.~\ref{fig:gan}). 
To leverage the knowledge beneath real images $\mathcal{D}_{s}$, $\mathcal{D}_{t}$ and synthetic ones $\mathcal{D}_{s}^s$, $\mathcal{D}_{t}^s$, we propose an MT framework for label-efficient UDA, named \ourmethod, in which, the student model explore the knowledge beneath \emph{source-like domain} and \emph{target-like domain} through two teacher models simultaneously for comprehensive integration.

\subsection{Dual Cycle Alignment Module}

\begin{figure}[htb]
    \centering
    \includegraphics[width = 0.9\columnwidth]{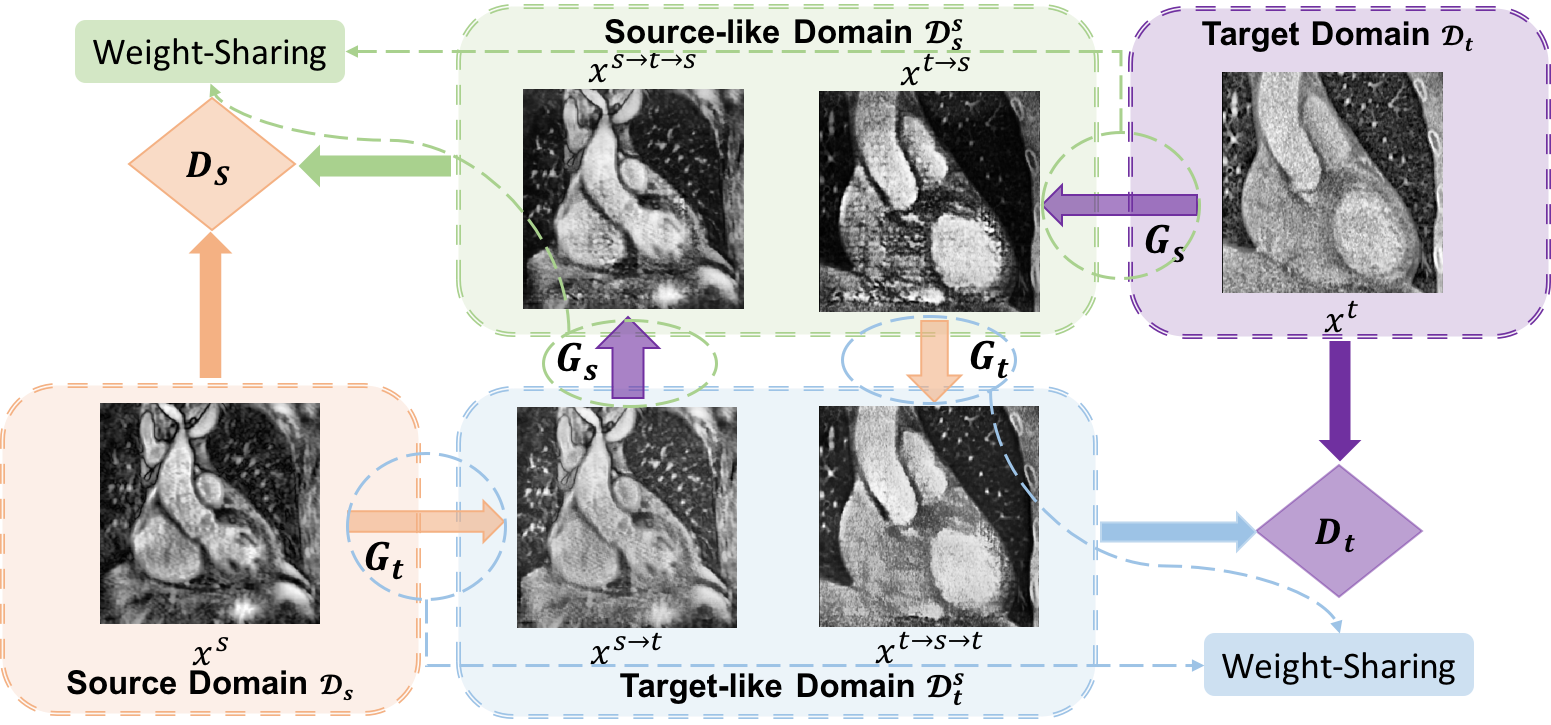}
    \caption{Overall framework of Dual Cycle Alignment Module (DCAM).}
    \label{fig:gan}
\end{figure}

To reduce the semantic gap across domains, we generate synthetic samples for two domains using generative adversarial networks~\cite{GAN}. 
We design a dual cycle alignment module (DCAM) based on CycleGANs~\cite{zhu2017unpaired} to narrow the domain shift bidirectionally, as demonstrated in Fig.~\ref{fig:gan}. 
To be specific, the target generator $G_{t}$ aims to transform source domain inputs to target domain distribution, \emph{i.e.}, $G_{t}\left(x^{s}\right)=x^{s \rightarrow t}$, whereas the discriminator $D_{t}$ aims to differentiate whether the images are fake target images $x^{s \rightarrow t}$ or real ones $x^{t}$. 
Similarly, with $x^{t}$, $G_{s}$ aims to generate $x^{t \rightarrow s}$, while $D_{s}$ aims to classify the transferred images $x^{t \rightarrow s}$ and the original images $x^{s}$. In CycleGAN, a reverse generator is employed to impose a cycle consistency between source domain images $x^{s}$ and reconstructed images $x^{s \rightarrow t \rightarrow s}$. 
It is noted that both the reverse generator and the source generator $G_{s}$ aim to generate source-like images, therefore, we share the weights between them. 
In similar fashion, we refactor $G_{t}$ to generate $x^{t \rightarrow s \rightarrow t}$. 
Different from CycleGAN, we further force the discriminator $D_{s}$ to differentiate source images $x^{s}$, synthetic source images $x^{t \rightarrow s}$ or reconstructed source images $x^{s \rightarrow t \rightarrow s}$ in order to bridge the domain gap better, 
Similarly, we construct a powerful discriminator $D_t$. 
Finally, we can obtain two newly-augmented intermediate domains, \emph{i.e.}, \emph{source-like domain} $\mathcal{D}_{s}^s = \{ x^{t \rightarrow s}, x^{s \rightarrow t \rightarrow s} \}$ and \emph{target-like domain} $\mathcal{D}_{t}^s = \{ x^{s \rightarrow t}, x^{t \rightarrow s \rightarrow t} \}$.

\subsection{Semantic Knowledge Transfer}

Following image-level adaptation by DCAM, \emph{source-like domain} images $\mathcal{D}_{s}^s$ and source domain images $\mathcal{D}_{s}$ maintain a similar visual appearance, allowing us to leverage the knowledge beneath $\mathcal{D}_{s}^s$ to improve the segmentation performance on $\mathcal{D}_{s}$ under label scarcity. 
As shown in Fig.~\ref{fig:pipeline}, we follow the mean teacher (MT) paradigm and adopt the same architecture for the student and teacher models based on self-ensembling~\cite{tarvainen2017mean}. 
Specifically, the teacher model $f_{\theta^{\prime}}$ at training step $t$ is updated with the exponential moving average (EMA) weights of the student model $f_{\theta}$, \emph{i.e.}, $\theta_{t}^{\prime}=\alpha \theta_{t-1}^{\prime}+(1-\alpha) \theta_{t}$, where $\alpha$ is the EMA decay rate that reflects the influence level of the current student model parameters. 
Given different perturbations (\emph{e.g.}, noises $\xi$ and $\xi^{\prime}$) to the inputs of teacher and student models, we expect their predictions to be consistent by minimizing the difference between them with a mean square error (MSE) loss $\mathcal{L}_{{tea }}^{{kd}}$ as 
\begin{equation}
\label{eq:loss_kd}
\mathcal{L}_{{tea }}^{k d}=\frac{1}{N} \sum_{i=1}^{N}\left\|f\left(x_{i} ; \theta_t^{\prime}, \xi^{\prime}\right)-f\left(x_{i} ; \theta_t, \xi\right)\right\|^{2},
\end{equation}
where $f(\cdot)$ is the segmentation network. $f\left(x_{i} ; \theta_t, \xi\right)$ and $f\left(x_{i} ; \theta_{t}, \xi^{\prime}\right)$ represent the outputs of the student model and the teacher model, respectively.

\subsection{Structural Knowledge Transfer}
Despite distinct differences like image appearance across domains, the transformed images obtained from generators should have the same structural information as the original ones. 
In other words, source domain image $x_{i}^s$ and its synthesis target-like image $x_{i}^{s\rightarrow t}$ should have the same segmentation masks, \emph{i.e.}, $ y^s= y_{i}^{s\rightarrow t}$.
In this regard, We propose a teacher model for keeping structural consistency between predictions of source images and corresponding synergistic target images, \emph{i.e.}, $f(x; \theta, \xi )=f(G_i [x] ; \theta^{\prime}, \xi^{\prime})$, where $x$ are source (-like) domain images, and $G_i$ is generator $G_{t}$ or reverse generator $G_{s}$. 
Transferring structural knowledge across domains not only regularizes the student model for semi-supervised learning, but also helps increase adaptation performance at the feature level.
Instead of the conventional consistency loss, \emph{e.g.}, MSE loss~\cite{li2020transformation}, we exploit the structural information based on weighted self-information~\cite{vu2019advent,vu2019dada}, and calculate the structural consistency loss $\mathcal{L}_{{tea }}^{{con}}$ between the teacher and student networks as
\begin{equation}
\label{eq:loss_con}
\mathcal{L}_{{tea }}^{{con}}=\frac{1}{N} \sum_{i=1}^{N} \frac{1}{H \times W} \sum_{v=1}^{\mathcal{V}}\left\|\mathbf{I}_{i, v}^{s}-\mathbf{I}_{i, v}^{t}\right\|^{2}
\end{equation}
where $\mathcal{V}=\{1,2, \ldots, H \times W \}$, $\mathbf{I}_{i, v}^{s} = -\mathbf{p}_{i, v}^{s} \circ \log \mathbf{p}_{i, v}^{s}$ is the weighted self-information of the predicted label at $v$-th pixel of $i$-th input from the student network, and similarly $\mathbf{I}_{i, v}^{t}$ is that from the teacher network. The notation $\circ$ is Hadamard product and log is the logarithmic expression using base $2$.

\subsection{MT-UDA Framework}
With the supervision of corresponding labels $y^{s}$, the student model is trained by the supervised loss $\mathcal{L}_{{stu }}^{s e g}$ as 

\begin{equation}
\label{eq:loss_stu}
\mathcal{L}_{{stu }}^{s e g}=\frac{1}{2}\left[\mathcal{L}_{ {ce }}\left(y^{s}, p_{{stu }}^{s}\right)+\mathcal{L}_{ {dice }}\left(y^{s}, p_{ {stu }}^{s}\right)\right],
\end{equation}

where $\mathcal{L}_{ {ce }}$ and $\mathcal{L}_{ {dice }}$ are cross-entropy loss and dice loss, respectively, and $p_{ {stu }}^{s}$ is the predictions of the student model on source labeled images $x^s$. Based on the above discussion, we integrate Eq.~\ref{eq:loss_kd}, Eq.~\ref{eq:loss_con} and Eq.~\ref{eq:loss_stu}, and the training objective for the student model is formulated as
\begin{equation}
\mathcal{L}_{s t u}=\mathcal{L}_{s t u}^{s e g}+\lambda_{k d} \mathcal{L}_{t e a}^{k d}+\lambda_{c o n} \mathcal{L}_{t e a}^{c o n}, 
\end{equation}

where $\lambda_{k d}$ and $\lambda_{c o n}$ are the trade-off parameters with the associated losses. With the \ourmethod~framework, we can distill the knowledge from \emph{source-like domain} and \emph{target-like domain} together for more accurate cross-modality image segmentation.

\section{Experiments and Results}

\subsubsection{Dataset and pre-processing.}
We evaluated our method on the Multi-Modality Whole Heart Segmentation (MM-WHS) 2017 dataset, consisting of unpaired $20$ MR and $20$ CT volumes with ground truth masks. We employed MR as source domain and CT as target domain. 
Following general UDA setting as in~\cite{chen2019synergistic}, each modality was first randomly split with $16$ scans for training and $4$ scans for testing. To validate the performance under the source-label scarcity scenario, we randomly selected $4$ annotated MR scans for training in comparison experiments.
For data pre-processing, following previous work~\cite{dou2018pnp}, we cropped all the coronal slices into centering at the heart region after resampling with unit spacing. Four cardiac substructures, \emph{i.e.}, ascending aorta (AA), left atrium blood cavity (LAC), left ventricle blood cavity (LVC), and myocardium of the left ventricle (MYO) were selected for segmentation.

\begin{table}
\centering
\caption{Comparison results of different methods. suffix -4 or -16 after method names stand for the number of labelled source scans used for training.}
\label{tab:results}
\begin{tabular}{c|c|c|c|c|c|c|c|c|c|c|c} 
\toprule
\multicolumn{2}{c|}{\multirow{2}{*}{Method}}                                                        & \multicolumn{5}{c|}{Dice~$\uparrow$ }                              & \multicolumn{5}{c}{ASD~$\downarrow$ }                          \\ 
\cline{3-12}
\multicolumn{2}{c|}{}                                                                               & AA             & LAC            & LVC            & MYO            & Avg            & AA            & LAC           & LVC           & MYO           & Avg            \\ 
\hline
\multicolumn{2}{c|}{W/o Adaptation - 4}                                                             & 5.6            & 17.8           & 12.1           & 5.5            & 10.3           & 36.9          & 24.6          & 38.5          & 35.6          & 33.9           \\ 
\hline
\multirow{3}{*}{UDA-16}                              & PnP-AdaNet~\cite{dou2018pnp} & 74             & 68.9           & 61.9           & 50.8           & 63.9           & 12.8          & 6.3           & 17.4          & 14.7          & 12.8           \\
                                                                      & SIFA-v1~\cite{chen2019synergistic}    & 81.1           & 76.4           & 75.7           & 58.7           & 73             & 10.6          & 7.4           & 6.7           & 7.8           & 8.1            \\
                                                                      & SIFA-v2~\cite{chen2020unsupervised}    & 81.3           & 79.5           & 73.8           & 61.6           & 74.1           & 7.9           & 6.2           & 5.5           & 8.5           & 7              \\ 
\hline
\multirow{2}{*}{UDA-4}                                                & DCAM                        & 19.3           & 28.1           & 34.1           & 6.4            & 22             & 32.5          & 21.8          & 17.7          & 22.8          & 23.7           \\
                                                                      & SIFA-v2~\cite{chen2020unsupervised}    & 50.5           & 59.6           & 31.9           & 28.9           & 42.7           & 8.8           & 7.3           & 15.8          & 13.2          & 11.3           \\ 
\hline
\multirow{2}{*}{SSL-4}                                                & MT~\cite{meanteacher}         & 3.6            & 26.8           & 14.5           & 4.6            & 12.4           & 34.5          & 22.7          & \textbf{5.7} & 17.6          & 20.1           \\
                                                                      & UA-MT~\cite{yu2019uncertainty}      & 20.1           & 40.5           & 2.5            & 11.3           & 18.6           & 40.1          & 23.3          & 43.2          & 20.9          & 31.9           \\ 
\hline
\multirow{3}{*}{\begin{tabular}[c]{@{}c@{}}UDA\\+SSL-4 \end{tabular}} & DCAM+MT~\cite{meanteacher}    & 35.3           & 31.6           & 48.4           & 11.2           & 31.6           & 39.9          & 39.8          & 10.5          & 14.6          & 23.7           \\
                                                                      & DCAM+UA-MT~\cite{yu2019uncertainty} & 61.3           & 59.7           & 46.5           & 19.2           & 46.7           & 5.6           & 8.3           & 8.2           & 10.6          & 8.2            \\
                                                                      & MT-UDA (Ours)               & \textbf{72.7}  & \textbf{71.4}  & \textbf{60.7}  & \textbf{41.7}  & \textbf{61.6}  & \textbf{5.3}  & \textbf{5.7}  & 6.7           & \textbf{6.1}  & \textbf{5.9}   \\
\bottomrule
\end{tabular}
\end{table}

\subsubsection{Implementation details.}

We followed~\cite{zhu2017unpaired} to optimize the proposed dual cycle alignment module for generating \emph{source-like domain} images $\mathcal{D}_{s}^s$  and \emph{target-like domain} images $\mathcal{D}_{t}^s$. 
Similar to~\cite{zhang2018task}, we verified our model on the transformed source-like images $x^{t \rightarrow s}$ instead of target domain images $x^{t}$, since our model was trained on source domain under source label scarcity. 
We implemented U-Net~\cite{ronneberger2015u} as our network backbone for both student and teacher models in \ourmethod. 
We trained the framework for a total of $150$ iterations and used Adam optimizer with the initial learning rate of $1\times10^{-4}$, momentum of $0.9$, learning rate warm up over the first $20$ iterations, and cosine decay of the learning rate with the SGD optimizer. 
Following~\cite{meanteacher}, the EMA decay rate $\alpha$ was set to $0.999$ for two teacher models, and hyperparameters $\lambda_{con}$ and $\lambda_{kd}$ were ramped up individually with the sigmoid-shaped function $\lambda(t)= 0.01\cdot e^{\left(-5\left(1-t / t_{\max }\right)^{2}\right)}$, where $t$ and $t_{max}$ were the current and the last step, respectively. 
Data augmentation such as random rotation was applied in all the experiments for a fair comparison. We evaluate different methods on Dice score and average surface distance (ASD) with the largest 3D connected component of each substructure.

\begin{figure}[t]
    \centering
    \includegraphics[width =0.85 \columnwidth]{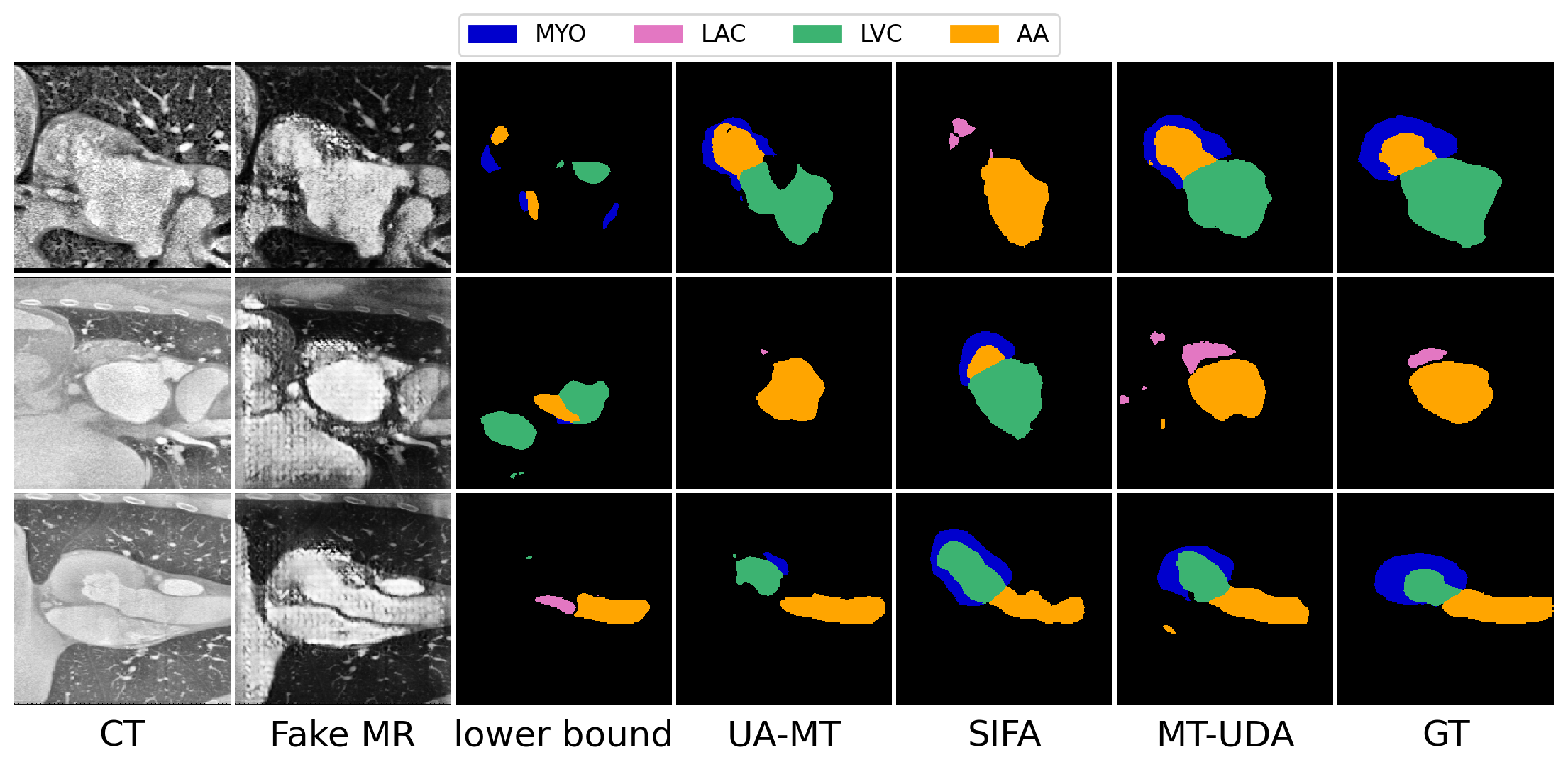}
    \caption{Visualization of segmentation results generated by different methods.}
    \label{fig:visual}
\end{figure}

\subsubsection{Comparison with other methods.} We compare our methods with the state-of-the-art UDA methods in cardiac segmentation, \emph{i.e.,} Pnp-AdaNet~\cite{pnp_1} and SIFA~\cite{chen2019synergistic,chen2020unsupervised}, as well as two recent popular SSL approaches, including MT~\cite{meanteacher} and UA-MT~\cite{yu2019uncertainty}. 
In Table~\ref{tab:results}, we list the results of PnP-AdaNet and SIFA with $16$ labeled source scans in cardiac segmentation. 
Since SIFA-v2~\cite{chen2020unsupervised} obtains the best segmentation performance on each substructure, we further train SIFA-v2 on $4$ labeled MR scans to simulate the source-label scarcity scenario. 
It is observed that SIFA-v2 obtains severely degraded performance on target domain when using $4$ labeled source domain scans, which can be attributed to the source label scarcity. 
We also directly test the U-Net trained on $4$ labeled MR scans from the source domain as our lower bound, referred as W/o Adaptation-4. 
By taking advantage of image-to-image translation \emph{i.e.,} DCAM, a great improvement can be achieved when testing W/o Adaptation on fake MR images $x^{t\rightarrow s}$, but it is still not optimal with the average dice of merely $22\%$ across the substructures. 
It is worth noting that MT and UA-MT can help improve the segmentation performance on target domain by leveraging unlabeled source domain images. 
Along with image appearance alignment, MT and UA-MT can achieve promising improvement on cross-modality segmentation, which demonstrates the feasibility of integrating SSL into UDA for label-efficient UDA. 
By simultaneously exploiting all available data sources, the proposed MT-UDA obtains the best segmentation results with the average dice of $61.6\%$, outperforming SIFA-v2 ($4$ training MR scans) by a large margin and achieving comparable performance with the state-of-the-art methods, but only requires $1/4$ source labels. 
We further visualize the segmentation results on testing data of different methods including the best methods of UDA and SSL, \emph{i.e.} SIFA-v2 and DCAM+UA-MT in Fig.~\ref{fig:visual}. It is observed that our method can generate more reliable masks with fewer false positives than other methods.

\begin{figure}[t]
    \centering
    \includegraphics[width =0.88 \columnwidth]{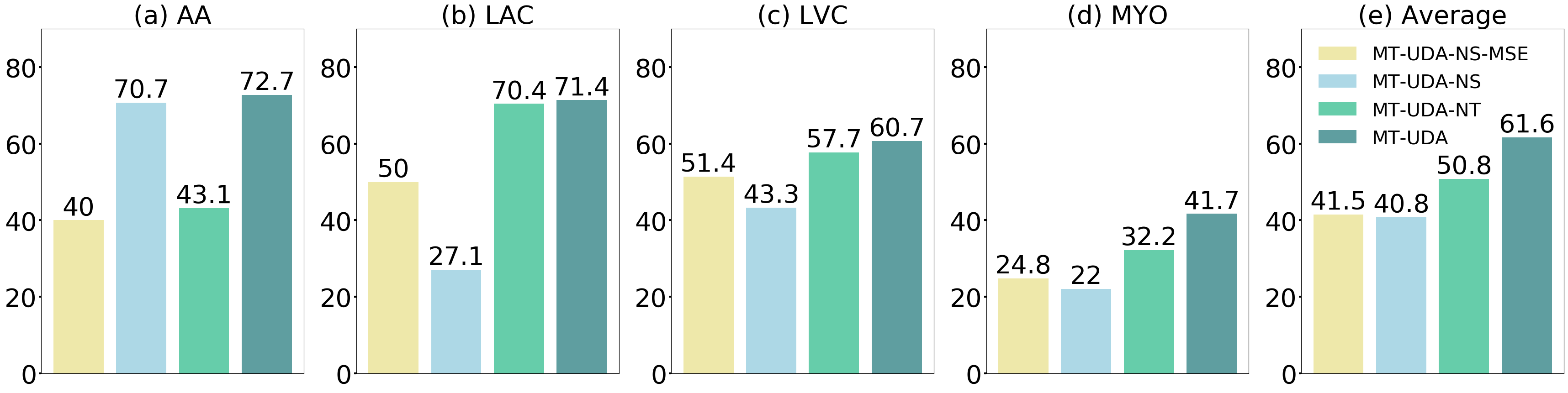}
    \caption{Ablation results (Dice[\%]) on different components.}
    \label{fig:abolation}
\end{figure}

\noindent\textbf{Ablation studies of our method.} 
To evaluate the effectiveness of different components of~\ourmethod, we conduct ablation experiments on various variants.
Specifically, we remove one of the teacher models, separately, \emph{i.e.}, W/o semantic knowledge transfer (MT-UDA-NS) and W/o structural knowledge transfer (MT-UDA-NT). 
We further implement the MSE loss in MT-UDA-NS to evaluate the efficacy of the structural loss, \emph{i.e.}, MT-UDA-NS-MSE. 
Fig~\ref{fig:abolation} demonstrates the ablation results of different substitutes. 
We can see that both types of knowledge transfer can benefit the model performance on unsupervised cross-domain segmentation. 
In comparison with the MSE loss in structural knowledge transfer, the proposed loss based on the weighted self-information can better improve the segmentation performance on some substructures such as AA, benefiting from the structural consistency across domains.

\section{Conclusion}
In this work, we present a novel label-efficient UDA framework, \ourmethod, which integrates SSL into UDA for cross-modality medical image segmentation under source label scarcity. 
By bridging both source and target domains to intermediate domains through knowledge transfer, the student model can leverage intra-domain semantic knowledge and exploit inter-domain structural knowledge concurrently, thereby mitigating both the domain discrepancy and source label scarcity. 
We evaluate the proposed \ourmethod~on MM-WHS 2017 dataset, and demonstrate that our method outperforms the state-of-the-art UDA methods by a lot under the challenging source-label scarcity scenario.

\subsubsection{Acknowledgement.}
This research is supported by Institute for Infocomm Research (I2R), Agency for Science, Technology and Research (A*STAR), Singapore.

%
%
%
\bibliographystyle{splncs04}
\bibliography{mybibliography}
%




\end{document}